\documentclass[preprint,aps,prd,12pt,preprintnumbers,eqsecnum,nofootinbib]{revtex4-2}
\usepackage{xcolor}
\usepackage{braket}
\usepackage{amssymb,amsmath,amsfonts,comment,latexsym,graphicx,epstopdf,float}
\usepackage{color}
\usepackage{fancyvrb}
\usepackage{chngcntr}
\usepackage{etoolbox}
\usepackage{ulem}
\usepackage{array}
\usepackage{bbold}
\usepackage[labelformat=empty]{subfig}
\usepackage{slashed}
\usepackage{multirow}

\usepackage[colorlinks=true,citecolor=blue,linkcolor=red,filecolor=cyan,urlcolor=magenta]{hyperref}

\usepackage{float}

\newcommand{\be}{\begin{equation}}
\newcommand{\ee}{\end{equation}}
\newcommand{\ba}{\begin{eqnarray}}
\newcommand{\ea}{\end{eqnarray}}

\newcommand{\grts}{\raise.3ex\hbox{$>$\kern-.75em\lower1ex\hbox{$\sim$}}}
\newcommand{\lets}{\raise.3ex\hbox{$<$\kern-.75em\lower1ex\hbox{$\sim$}}}

{\catcode`\|=\active\gdef\Braket#1{\left<\mathcode`\|"8000\let|\bravert 
{#1}\right>}}

\def\bravert{\egroup\,\vrule\,\bgroup}

\usepackage{color}
\usepackage{amssymb,amsmath,amsfonts}
\usepackage{epstopdf} 
\usepackage{braket}

\usepackage{autobreak}

\begin{document}
%
%  
% Title of paper
\title{\vspace*{0.5in} 
Towards a quantum field theory description of nonlocal spacetime defects 
\vskip 0.1in}
\author{Christopher D. Carone}\email[]{cdcaro@wm.edu}
\author{Noah L. Donald }\email[]{nldonald@wm.edu}
%`
\affiliation{High Energy Theory Group, Department of Physics,
William \& Mary, Williamsburg, VA 23187-8795, USA} 
%
%  
%\date{\today}
\date{October 6, 2023}
\begin{abstract}
We propose an ansatz for encoding the physics of nonlocal spacetime defects in the Green's functions for a scalar field theory
defined on a causal set.   This allows us to numerically study the effects of nonlocal spacetime defects on the discrete Feynman 
propagator of the theory defined on the causal set in 1+1 dimensions, and to compare to the defect-free limit.  The latter approaches 
the expected continuum result, on average,  when the number of points becomes large. When defects are present, two points with the 
same invariant spacetime interval can have different propagation amplitudes, depending on whether the propagation is between two 
ordinary spacetime points, two defects, or a defect and an ordinary point.  We show that a coarse-grained description that is only sensitive 
to the average effect of the defects can be interpreted as a defect-induced mass and wave-function renormalization of the scalar theory.
\end{abstract}
\pacs{}

\maketitle
\newpage
\section{Introduction} \label{sec:intro}

Gravitational physics in four dimensions is characterized by the Planck scale, $M_\text{Pl} \approx 1.2 \times 10^{19}$~GeV, where 
Newton's gravitational constant is given by $G_N = 1/ M_\text{Pl}^2$, in units where $\hbar=c=1$.  From a phenomenological perspective, 
this energy scale is frustratingly high.  The effects of physical states with Planck-scale masses, particularly in the absence of the violation of 
any fundamental symmetries, are far outside the reach of direct collider tests.   Cosmological or multi-messenger observations may reveal 
hints of quantum gravitational effects~\cite{Addazi:2021xuf}, though these approaches are still limited by the remoteness of the scale that is 
being probed.

It is therefore worthwhile to consider physics that may be implied by an underlying theory of quantum gravity, but that is not characterized 
by the same length scale.   Spacetime defects are such an example.   In models of quantum gravity that are based on the assumption that 
spacetime is discrete, it has been argued that spacetime defects are generic~\cite{Hossenfelder:2014hha,Hossenfelder:2013zda,Hossenfelder:2013yda}.   
Defects may be local~\cite{Hossenfelder:2013zda} or nonlocal~\cite{Hossenfelder:2013yda} depending on how a particle behaves when its world 
line intersects the location of a defect.   Nonlocal defects may translate an incident particle to another point on the spacetime manifold, potentially a 
macroscopic distance away.  For example, in modeling particle kinematics in the presence of nonlocal defects on a flat background, it was assumed 
in Ref.~\cite{Hossenfelder:2013yda} that relativistic particles are translated between nonlocal defects and emerge with their four-momenta 
unchanged;  both timelike and spacelike separations of the entry and exit points were considered and bounds on the spacetime density of 
nonlocal defects were obtained.

Here, we focus instead on a quantum field theory formulation of nonlocal defects.   We define defects, in general, as points on a spacetime 
manifold that are distinguished by their nontrivial interaction with, or effect on the propagation of, quantum fields.  The defect points are thus 
distinguished from the other points in the manifold, which we will assume form a discrete set.  Spacetime translation invariance is broken 
when taking the continuum limit of the ordinary ({\it i.e.,} non-defective) points on a locally flat patch that contains a defect.  The literature on 
quantum field theories in the presence of local spacetime defects is small~\cite{Schreck:2012pf,Klinkhamer:2017nhl,Queiruga:2017nfe}, while that of nonlocal 
defects is virtually nonexistent.   The aim of the present work is to take some preliminary steps towards formulating quantum field theories in the presence 
of nonlocal defects, a least in the setting of a toy scalar model in a two-dimensional discrete spacetime, where numerical calculations are tractable.

For modeling a discrete spacetime, we will adopt the well-known framework of causal sets~\cite{Bombelli:1987aa,Surya:2019ndm}.  We 
review this approach in the next section.  We assume some fraction of causal set elements are defect points (which we define precisely 
later) and that this fraction persists as the continuum is approached so that we do not end up with a situation where the probability of
encountering a defect becomes vanishingly small.   Crucially, our set up assumes two different densities of points that can be separately adjusted, which provides two different fundamental length scales.   The hope is that this will allow the nonlocality scale of the defects to be larger than the nonlocality that is expected with the causal set construction of ordinary spacetime.  We discuss this distinction later when we define our models. 

We will focus on the computation of the Feynman propagator for a real scalar field on the set of ordinary and defect points, using a 
Green's function approach to field quantization that was originally proposed by Johnston~\cite{Johnston:2008za,Johnston:2009fr}.  What 
is particularly convenient about this approach in the present context is that the Green's functions are derived by summing over the 
trajectories of particles over discrete spacetime points, even though the resulting discrete expression for the Green's functions are 
subsequently used in quantizing a scalar field theory.  Johnston showed that this approach reproduces the expectation for position-space 
Feynman propagator as the continuum limit is approached.  Our {\it ansatz} is to encode the physics of defects in the sum over particle 
trajectories that define the Green's functions; we then study the effect on the Feynman propagator that is obtained when those modified 
Green's functions are applied to the quantization of a scalar field on the causal set.

Our paper is organized as follows.  In the next section, we review the approach to computing the Feynman propagator for a real scalar field
on a causal set.  In Sec.~\ref{sec:models}, we  explain how we modify the Green's functions used in the scalar field quantization to take into account the presence of spacetime defects.  In Sec.~\ref{sec:results}, we present our numerical results for the models that we define, 
comparing to the limit where the density of spacetime defects is taken to zero.  In Sec.~\ref{sec:conc} we summarize our conclusions.

\section{Scalar propagator on a causal set} \label{sec:review}

Before reviewing the approach of Refs.~\cite{Johnston:2008za,Johnston:2009fr} to scalar field quantization, we first recall some basic definitions of causal set theory.  A causal set is a locally finite, 
partially ordered set, $(\mathcal{C},\preceq)$, satisfying the following properties for any $v_{i},v_{j},v_{k} \in \mathcal{C}$:
\begin{enumerate}
        \item Reflexive: $v_{i}\preceq v_{i}$.
        \item Transitive: If $v_{i}\preceq v_{j}$ and $v_{j}\preceq v_{k}$, then $v_{i}\preceq v_{k}$.
        \item Antisymmetric: If $v_{i}\preceq v_{j}$ and $v_{j}\preceq v_{i}$, then $v_{i}=v_{j}$.
        \item Locally Finite: If $v_{i}\preceq v_{k}$, then the set $\{v_{l} \in \mathcal{C} | v_{i}\preceq v_{l} \preceq v_{k}\}$ has finite cardinality.
\end{enumerate}
   
Physically, the elements of the causal set label spacetime events and the notation $v_{i} \prec v_{j}$ denotes that a spacetime event $v_{i}$ is in the 
causal past of an event $v_{j}$. Conditions 1-3 are needed to define a consistent causal structure and condition 4 ensures discreteness. 

A straightforward method of obtaining a causal set which approximates a Lorentzian manifold, called sprinkling, is to begin with the 
Lorentzian manifold and randomly select a set of discrete points such that the number in any closed region of finite volume $V$ is given
by a Poisson distribution, with mean $\rho \,V$.  The causal relations amongst the Poisson distributed points are then inherited from the 
Lorentzian manifold. 
   
To study particle propagation, suppose we have a finite causal set Poisson sprinkled at some density, $\rho$, in a 1+1 dimensional 
region of flat spacetime and we enumerate the events in the causal set. Recall the definition of the causal matrix: $C_{ij}=1$ if $v_{i} \prec v_{j}$ 
and $C_{ij}=0$ otherwise. Thus, the causal matrix encodes all of the causal structure of the causal set.  Define a chain of length $k$ to be a 
subset of the causal set containing $k$ elements which is totally ordered. In other words, all $k$ elements are causally related to each 
other. The causal matrix has the useful property that $[C^n]_{ij}$ is the number of chains of length $n$ from element $v_{i}$ to $v_{j}$ in the causal set. 

Lets explore these chains from the viewpoint of a particle propagating within the causal set. To do so, we will follow the approach of 
Refs.~\cite{Johnston:2008za,Johnston:2009fr} and assign an amplitude, $a$, for moving between two causally related elements as well as 
an amplitude, $b$, for stopping at an intermediary element in a chain. Thus, the total amplitude for moving between two elements of the 
causal set in $k$-steps is $b^{k-1}a^{k} \, C^{k}$.\footnote{For an alternative formulation, see Ref.~\cite{Shuman:2023agb}}  If we define 
$\phi=a \, C$, then this amplitude can be expressed as $b^{k-1}\phi^{k}$. If we want to consider chains of all lengths, then we need to 
perform the sum. Note that for finite causal sets, there exists a positive integer, $\ell$, such that $[\,C^{\ell\,}]_{ij}=0$. Thus, the sum over chains of 
all possible lengths is a finite sum. If chains of length zero are included, then, the total amplitude, $K$, is:\footnote{It may be better to think of the product $a b$ as a
weighting factor associated with steps taken between causal set points, and $1/b$ an overall weighting factor for the sum over non-zero length paths.  This avoids
the necessity of explaining what it means to move to a point without stopping.}
\begin{equation}
       K=I + \sum_{i=1}^{\infty}b^{i-1}\phi^{i}=I+\phi \, [I-b\, \phi]^{-1}. 
\end{equation}
From this, Johnston then defines, by analogy with the continuum, the retarded propagator, $K_{R}$, the advanced propagator, $K_{A}$, and the Pauli-Jordan function, $\Delta$, in this discretized setting~\cite{Johnston:2009fr}:
\begin{equation}
K_{R}=\phi \, [I-b\, \phi]^{-1} \, ,
\label{eq:addup}
\end{equation}
\begin{equation}
K_{A}=K_{R}^{T} \,\,\, ,
\end{equation}
\begin{equation}
\Delta=K_{R}-K_{A} \,\,\, .
\label{eq:csdelta}
\end{equation}
The matrix $i\Delta$ is both hermitian and skew-symmetric which guarantees that its eigenvalues come in positive and negative pairs.  One 
can project $i\Delta$ onto its positive eigenvalue subspace by performing the sum: 
\begin{equation}
Q=\sum_{\, i,\,\lambda_i>0}\lambda_i \, u_i \, u_i^{\dag}.   
\end{equation}  
Here, the index $i$ labels elements in the spectrum of $i\Delta$ and $u_i$ is the eigenvector corresponding to the eigenvalue $\lambda_i$.

In the continuum quantum field theory, canonical quantization implies the commutator $[ \phi(x) , \phi(y)] = i \, \Delta(y-x)$; by analogy, 
one requires
\begin{equation}
[ \phi_x , \phi_y] = i \Delta_{xy} 
\end{equation}
in the causal set theory, with the right-hand-side following from Eq.~(\ref{eq:csdelta}).  This connects the Green's functions defined on the causal 
set to the algebra of scalar field operators.  It is shown in Ref.~\cite{Johnston:2009fr}  that this algebra can be used to derive an expression for the 
Feynman propagator, a vacuum-to-vacuum matrix element of the time-ordered product of fields
\begin{equation}
(K_F)_{xy} = i \, \langle 0 |\,  T \, \phi_x \, \phi_y \, | 0 \rangle \,\,\, .
\end{equation}
Here, the factor of $i$ follows the convention of Ref.~\cite{Johnston:2009fr}, where one can also find the definition of the vacuum states.  The 
real and imaginary parts of $K_F$ can be expressed in terms of the quantities already defined:
\begin{equation}
{\rm Im}(K_{F})={\rm Re}(Q) \,\, ,
\label{eq:realK}
\end{equation}
\begin{equation}
{\rm Re}(K_{F})=\frac{1}{2}(K_{R}+K_{A}).
\label{eq:imK}
\end{equation}
Averaging over possible Poisson sprinklings,  one recovers the position space continuum Feynman propagator for a real 
scalar field with mass $M$, provided one chooses $a=1/2$ and $b=-M^2 / \rho$ in 1+1 
dimensions~\cite{Johnston:2008za,Johnston:2009fr}.  For the purposes of comparison, the continuum position-space Feynman 
propagator in 1+1 dimensions is given by
\begin{equation}
G_{F}(x)=\frac{1}{4}H_{0}^{(2)}(Ms) \,\,\, ,
\end{equation}
where $H_{0}^{(2)}$ is the zeroth order Hankel function of the second kind. Here, $s=\sqrt{(x^{0})^2-(x^{1})^2}$ when $(x^{0})^2 \geq (x^{1})^2$, and $s=-i\sqrt{(x^{1})^2-(x^{0})^2}$ 
when $(x^{0})^2 \leq (x^{1})^2$.    

The way we identify defects in the construction that we have just reviewed is twofold: (1) we may assign a weighting factor for moving to or from defect points that differs from the value of $a$ that is 
associated with hopping between ordinary spacetime points, and (2) we may restrict movement to or from defect points.   For example, in the first model that we consider in the next section, a particle 
encountering a defect may not be allowed to propagate to any spacetime point in the forward light cone, but only to one specified point.  We will refer to this type of propagation as ``beaming," to distinguish it from 
the unrestricted case.  We will also separately Poisson sprinkle ordinary and defect points, so that we can control their spacetime density separately.  This gives us two fundamental length scales that can be separately adjusted while maintaining the Lorentz invariance of the theory~\cite{Dowker:2003hb,Bombelli:2006nm}.

\section{Models of Defects} \label{sec:models}
The expressions for the Feynman propagator in discrete spacetime, Eqs.~(\ref{eq:realK}) and (\ref{eq:imK}), depend on the causal matrix $C$, as well as the constants
$a$ and $b$.  The models we consider for defects are encoded in the detailed form of $C$, with the values of $a$ and $b$ chosen so that the model
reproduces the expected results in the absence of defects.  We consider five scenarios, that we label A though E, which we study in
two-dimensional Minkowski spacetime.  Raising the modified causal matrix to a suitably high power in each model below will return the zero matrix, the property assumed in deriving Eq.~(\ref{eq:addup}).

\begin{figure}[t]
\centering
\includegraphics[width=0.45\textwidth]{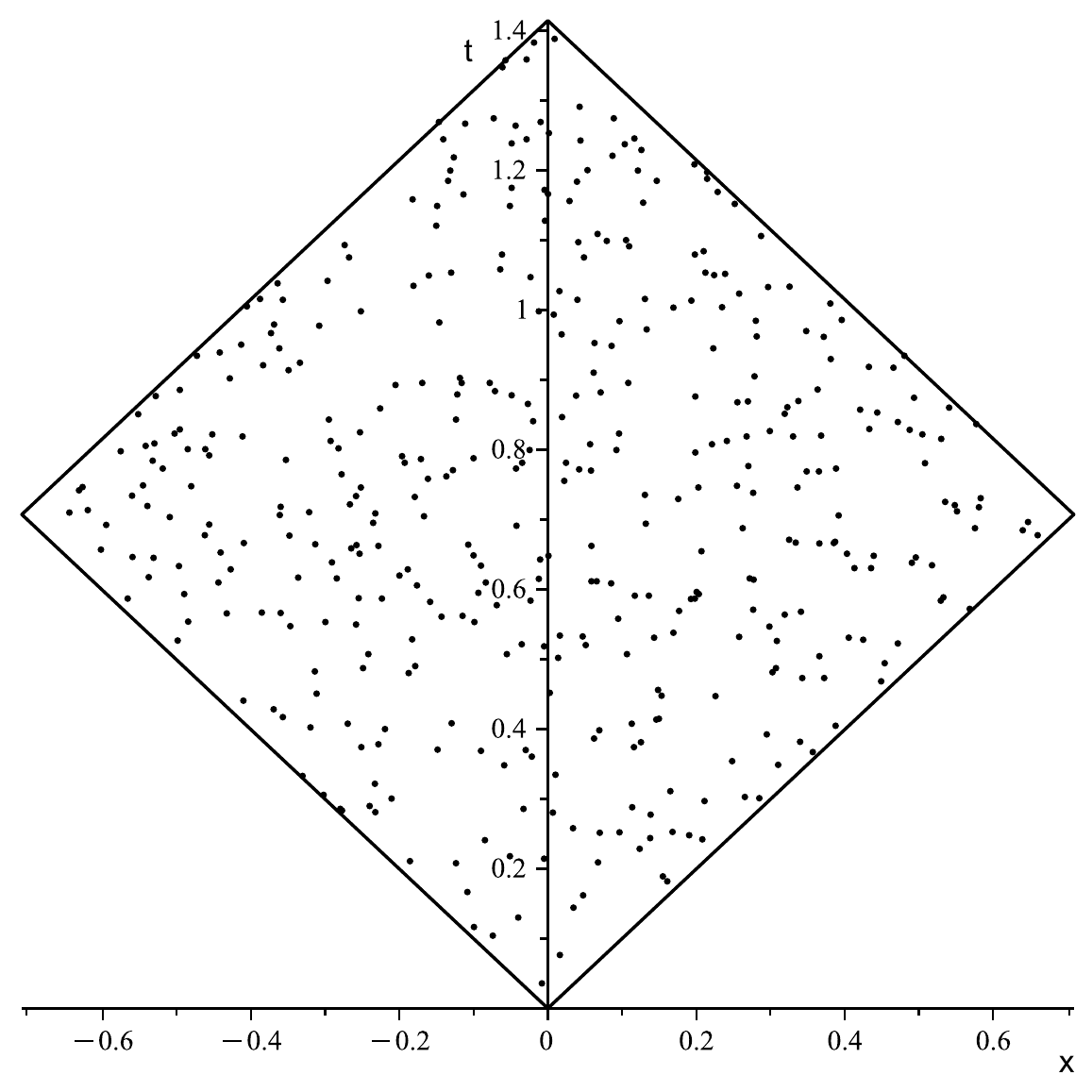}
\includegraphics[width=0.45\textwidth]{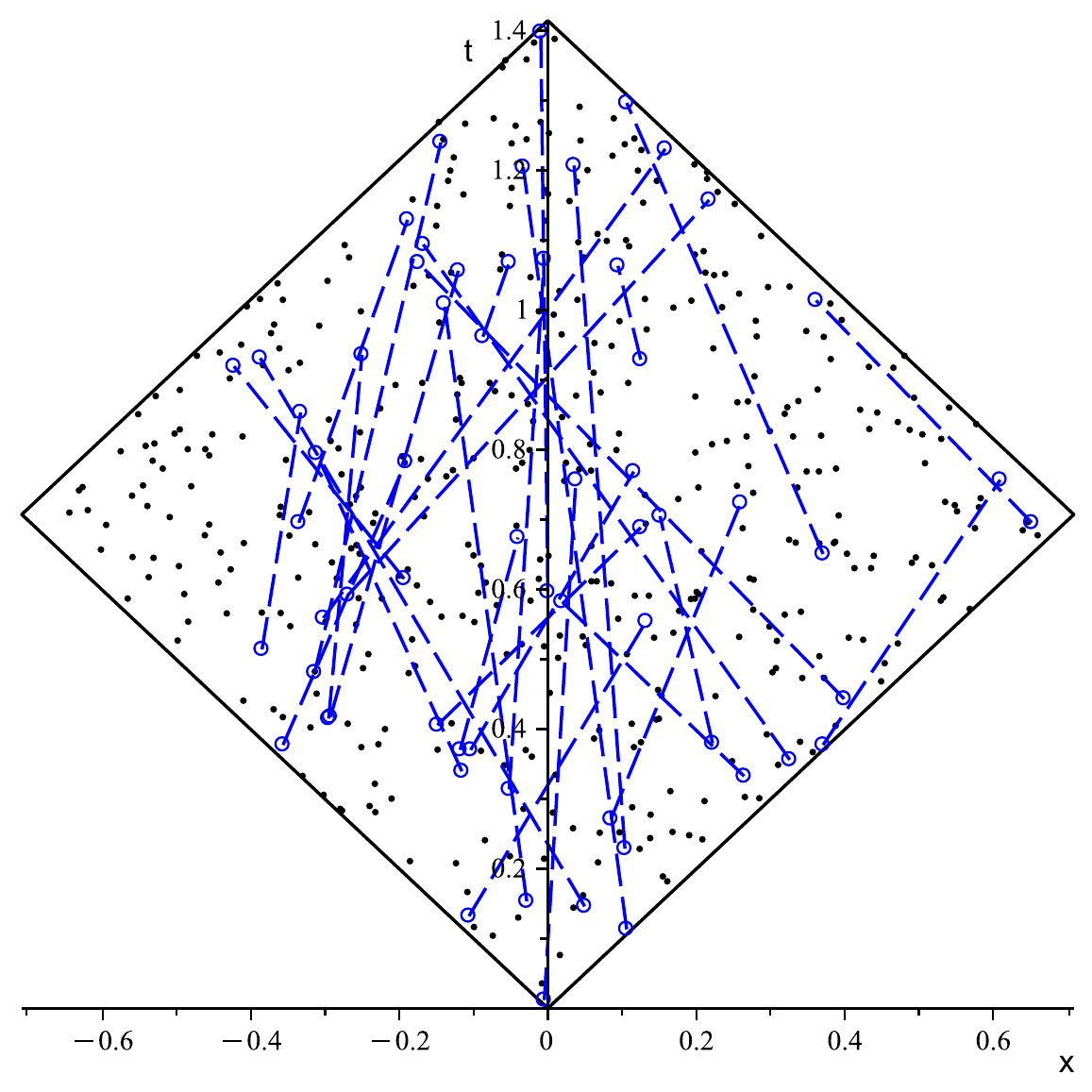}
\caption{{\it Left}: Poisson sprinkling of $400$ spacetime points in two dimensions, in unit spacetime volume.  {\it Right}: The same with the addition of 
$25$ pairs of linked, ``in" and ``out" defects, as in Model A, indicated by timelike dashed line segments.}
\label{fig:sprinkles} 
\end{figure}

{\bf Model A}.  Conceptually the simplest scenario we can imagine is that we Poisson sprinkle two sets of defect points, for ``in" and ``out" defects.  In summing
over all possible particle trajectories, we assume that a trajectory incident on an ``in" defect is beamed exclusively to a specific ``out" defect.  
We implement the one-to-one pairing on the ``in" and ``out" defects randomly, with the ``out" defect always identified with the point in the pair with the
larger time coordinate.   This ordering will remain the same in any frame if we build our in-out sets so that all pairs have timelike separation, which is the case in 
Model~A.   One might think of these paired defects as the entrance and exit points of a wormhole, though that would only be a convenient metaphor; there are no 
spacetime points traversed between these points and nothing analogous to a nontrivial spacetime topology between them.   To give a pictorial representation of 
this model, we show in Fig.~\ref{fig:sprinkles}, the Poisson sprinkling of 400 spacetime points in a causal diamond of unit volume, compared to the same with an additional 
sprinkling of 25 ``in" defects and 25 ``out" defects that form timelike pairs, indicated by the dashed line segments.  This can be encoded in a causal matrix that has the following
blocks
\begin{equation}
C = \left(\begin{array} {cc|c|cc} c_{11} &  \hspace{1em} &  0 & \hspace{1em} & c_{13} \\ \hline
c_{21} &&  0 &&  c_{23} \\ \hline
0 && \epsilon/a \cdot \mathbb{1} && 0 \end{array} \right)  \,\,\, ,
\label{eq:CmodelA}
\end{equation}
where the basis we use corresponds to the set
\begin{equation}
\left\{p_1, p_2, \ldots p_N \, | \, O_1, O_2, \ldots, O_n \, | \,   I_1, I_2, \ldots, I_n \right\} \,\,\, .
\end{equation}
Here, the $p_j$ represent ordinary spacetime points, the $I_j$ are ``in" defect points and the $O_j$ are ``out" defect points, with the in-out pairing indicated
by the common value of the index $j$.  The sub-matrices $c_{ij}$ are causal matrices that are constructed in the usual way, but defined over the particular
subset of points.  For example, $c_{11}$ is the causal matrix that one constructs over the set of ordinary spacetime points ${p_1,\ldots p_N}$.  The block
proportional to the identity matrix enforces that particles encountering ``in" defects propagate only to a specific ``out" defect, potentially with a different weighting
factor in the sum over all trajectories, given by the parameter $\epsilon$.  

We assume the parameter values $a=1/2$ and $b=-M^2/\rho_T$, where $\rho_T$ is the total density of points and defects, 
$\rho_T = \rho+\rho_{\rm in}+\rho_{\rm out}$.  This assures that there is a limit where we approach the correct continuum results when the defects are absent, {\it i.e.}, when
$n \rightarrow 0$, so that $\rho_{\rm in} \rightarrow 0$ and $\rho_{\rm out} \rightarrow 0$.  We note that in this simple model, we require that $\epsilon \neq 0$; if this weren't the case, particles 
could propagate to an ``in" defect and then become trapped there.  We exclude this possibility as pathological.

{\bf Model B}.  This model is the same as Model A, with the same form for $C$ given in Eq.~(\ref{eq:CmodelA}), except that the paired ``in" and ``out" defects are 
spacelike separated.   Spacelike separated pairs of nonlocal defects were discussed in Ref.~\cite{Hossenfelder:2013yda}, though they may lead to a violation of 
causality without additional assumptions.  We include this case to illustrate the violation of causality in the quantum field theory context:  if the sum over particle 
trajectories includes these spacelike pairs, then the resulting scalar field theory will reveal that there is a problem if the commutator of fields for spacelike-separated points is nonvanishing.  This commutator is precisely the Pauli-Jordan function $\Delta$, which we evaluate numerically for Model B in the next section.

{\bf Model C}.  In Model A, a particle reaching an ``in" defect could only propagate to a specified ``out" defect, and no trajectories were included where an ``out" defect could be reached by any other route.  This presents a simple scenario, but one that might be overly restrictive.  For this reason, we consider a variation on Model A, where we allow the possibility that particles may reach and travel from ``in" and ``out" defects via 
any causal trajectory, but where travel between them is separately weighted. For example, a particle reaching an ``in" point may either beam preferentially to the exit point,  or propagate to any other point in the future light-cone, like it would at an ordinary spacetime point.   We call this Model C; its causal matrix is given by 
\begin{equation}
C = \left(\begin{array} {cc|c|cc} c_{11} &  \hspace{1em} &  c_{12} & \hspace{1em} & c_{13} \\ \hline
c_{21} &&  c_{22} &&  c_{23} \\ \hline
c_{31}  &&  c_{32} -(1-\epsilon/a) \mathbb{1}&& c_{33}
\end{array} \right). \,\,\, .
\label{eq:CmodelC}
\end{equation}
The $3$-$2$ block of this matrix is the usual causal matrix, with each diagonal entry modified from $1$ to $\epsilon/a$.  Again we assume $a=1/2$ and $b=-M^2/\rho_T$, where 
$\rho_T = \rho+\rho_{\rm in}+\rho_{\rm out}$.   We note that there are now two ways in which Model C can approach the no-defect limit:  either one may take $n$ to zero, so that 
there are just $N$ ordinary spacetime points with $\rho_T=\rho$, or one may take $\epsilon=a$, in which case one may compare to a theory with $N+2\, n$ ordinary spacetime points.

{\bf Model D}.  In this model, we assume $N$ ordinary points, with causal propagation to any other ordinary point, and $n$ defect points with causal propagation 
to any other defect point.  We allow these two sets to be coupled together causally, with the freedom to adjust the strength of the coupling ({\it i.e.}, how easy it is to enter the random lattice of defect points) and the weighting of the defect-to-defect propagation relative to the propagation between ordinary spacetime points.  These are controlled by the parameters $\xi$ and $\epsilon$, respectively, in the causal matrix for this scenario
\begin{equation}
C = \left(\begin{array} {c|c} c_{11} &    \xi/a \cdot c_{12}  \\ \hline
\xi/a \cdot c_{21} &  \epsilon/a \cdot c_{22} \end{array} \right) \,\,\, .
\label{eq:CmodelD}
\end{equation}
Here, the basis corresponds to the set
\begin{equation}
\left\{p_1, p_2, \ldots p_N \, | \, D_1, D_2, \ldots, D_n \right\} \,\,\, ,
\label{eq:basis2}
\end{equation}
where the $p_j$ are $N$ ordinary points and the $D_j$ represent $n$ defect points.  Rather than a model of defects, Eq.~(\ref{eq:CmodelD}) 
might be considered  as a way of modeling discrete nonlocality,  at potentially a larger scale than might be expected in a minimal causal set model 
of spacetime at the Planck scale.   Whatever the reader's preferred interpretation, this case is included in our numerical results.

It should be noted that the exact relationship between the fundamental scale set by the spacetime density of causal set points and the scale of 
nonlocality that appears in a low-energy effective field theory description is somewhat uncertain.  It was pointed out in Ref.~\cite{Sorkin:2007qi}, 
that a possible generalization of the d'Alembertian operator that emerges from a causal set description involves a free parameter that controls the
nonlocality of the low-energy theory.  It was argued that this scale should be lower than the fundamental scale from considerations of 
convergence, but that a large separation in scales would introduce a hierarchy problem.   We will assume these scales are not widely separated, 
based on naturalness; in any case, our construction introduces {\it two} independent scales that need not be similar.   It would be interesting to see 
how this maps into a low-energy continuum description, though that goes beyond the scope of the present work.

{\bf Model E}.  As a contrast to the other models that we have defined, we might interpret the following causal matrix as a model of local defects (at least as local as they can be in a model based on a causal set spacetime):
\begin{equation}
C = \left(\begin{array} {c|c} c_{11} &  c_{12}  \\ \hline
\kappa/a \cdot c_{21} &  \kappa/a \cdot c_{22} \end{array} \right) \,\,\, ,
\label{eq:CmodelE}
\end{equation}
again assuming the same basis as in Eq.~(\ref{eq:basis2}).  The adjustment in weighting given by the parameter $\kappa$ implies that if a defect point is reached, there is a different weighting for hopping away compared to hopping away from an ordinary spacetime point; there is no preferred coupling to a paired defect as in Models A, B or C.  Numerical results for the propagator in this case are also presented in the next section.

\section{Results} \label{sec:results}
\begin{figure}
\centering
\subfloat[\label{subfig:2a}]{%
  \includegraphics[width=0.5\columnwidth]{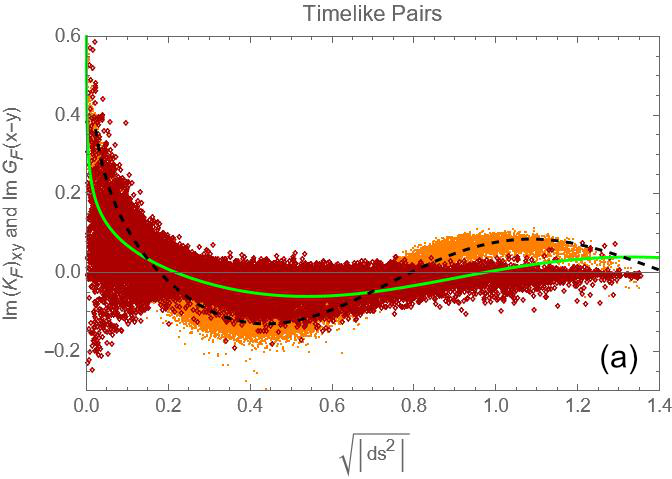}%
}\hfill
\subfloat[\label{subfig:2b}]{%
  \includegraphics[width=0.5\columnwidth]{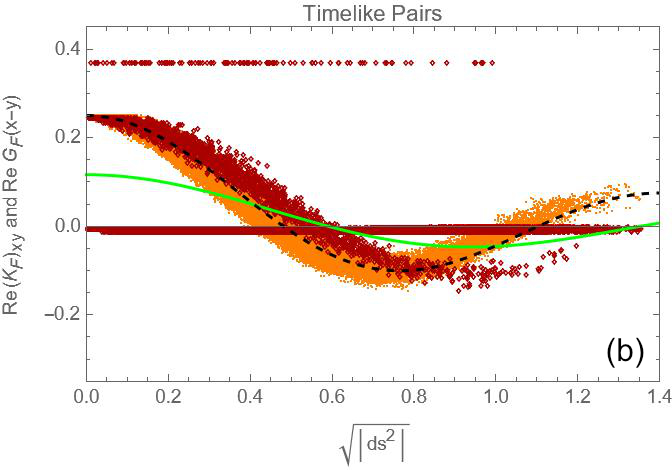}%
}\hfill
\subfloat[\label{subfig:2c}]{%
  \includegraphics[width=0.5\columnwidth]{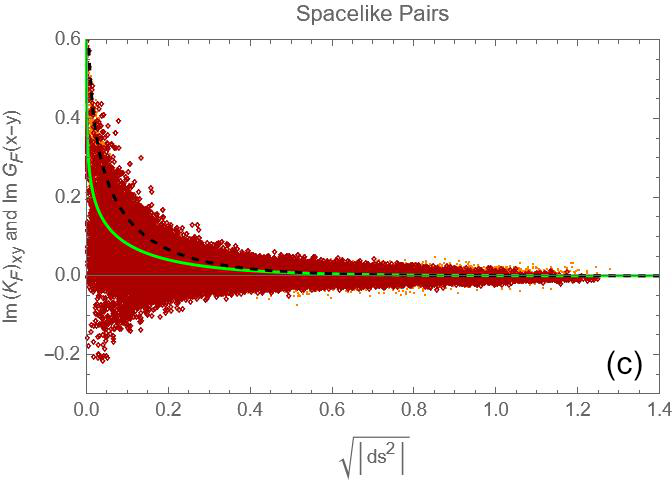}%
}
\subfloat[\label{subfig:2d}]{%
  \includegraphics[width=0.5\columnwidth]{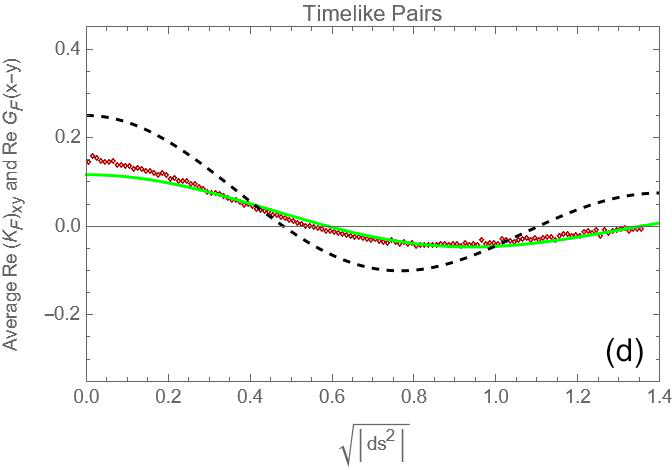}%
}
\caption{{\bf Model A} with $M=5$ and $\epsilon=0.75$.  Feynman propagator for a 600-element causal set that includes 200 defect 
points in timelike in-out pairs, in 1+1 dimensions.  (a) and (b) show the imaginary and real parts of the propagator, respectively, for 
points that are timelike separated; (c) shows the imaginary part for points that are spacelike separated (the real part vanishes).   The darker
points (red in color) are the results; the lighter points (orange in color) are the results in the absence of defects. The dashed curve (black in color) is 
the continuum propagator without defects and the solid curve (green in color) is the fitted curve to the average defective propagator; (d) is the 
same as (b) but only displays the average points in the presence of defects.
\label{fig:two}}
\end{figure}

We plot our results for the discrete Feynman propagator as a function of the absolute value of the proper time interval for pairs of 
elements in the causal set, as in Ref.~\cite{Johnston:2009fr}.  We work in units where the spacetime volume that contains the Poisson
sprinkling of points is unity.  When defects are present, a generic feature that we see when we plot 
the propagator as a function of the magnitude of the proper time interval is multivaluedness.  By this, we don't mean scatter about the curve associated with the 
defect-free limit, but convergence of some of the points to a distinct curve.  This feature could be anticipated: the value of the 
propagator between pairs of points with identical proper time intervals can nonetheless differ depending on whether the points in that 
pair are ordinary spacetime points, in-defects or out-defects (in the cases where there is a distinction).  This is guaranteed since the 
Green's functions used to define the theory are constructed assuming defect points have their own set of hopping amplitudes as well 
as restricted connections to other elements of the causal set.  In principle, one could imagine that an experiment with sufficient 
resolution might discern different results for a scattering amplitude depending, for example, on the location of the reconstructed 
production and decay vertices of an s-channel resonance, which would be sensitive this effect.  In other words, with sufficient 
resolution, an experiment might discern the breaking of translation invariance originating from the presence of the defects.   On the 
other hand, resolution might be insufficient.  (This is easy to imagine; consider Planck-scale discrete spacetime and a defect length 
scale that is, say, a thousand times larger.) In this case, one might instead work with a coarse-grained low-energy effective theory 
involving continuous quantities that are be defined by averaging over spacetime and defect points within a volume determined by 
the coarse-graining scale.  Averages of our results are also included in the examples that follows.
        
\subsection{Model A}
Model A introduced timelike in-out defect pairs in the sum over chains used in constructing the Green's functions for the theory.  
Figure~\ref{fig:two} displays the real and imaginary parts of $(K_F)_{xy}$, omitting the real part for spacelike separated points, which 
we find is vanishing.  The magnitude of the proper time interval between the points $v_x$ and $v_y$ is denoted $\sqrt{|ds^2|}$.   It is 
easiest to discern three curves, which we will refer to as branches of the solution, in Fig.~\ref{subfig:2b}.  Two are approximately 
constant, while one is similar to what is found in the no-defect case, though with a different shape.   The non-constant branch of 
$(K_{F})_{xy}$ gets its largest contribution when $x$ and $y$ are both ordinary spacetime points.  Since the density of spacetime 
points is greater than that of the defects, these propagator values give the dominant contribution when averaging the results at 
fixed $\sqrt{|ds^2|}$, shown as the solid curve in the figures.
        
For illustration, Fig.~\ref{subfig:2d} shows the same case as Fig.~\ref{subfig:2b}, but with points indicating the average result at a 
given value of  $\sqrt{|ds^2|}$.  The points roughly follow a curve that is of the same form as the position-space continuum propagator in a spacetime 
with no defects.  Hence, we fit these average points the same functional form, with two fit parameters
\begin{equation}
G_{F}(x)=\frac{Z}{4}H_{0}^{(2)}(M_r \, s) \,\,\, .
\label{eq:fitfnc}
\end{equation}
The first parameter, $Z$, may be identified as a wavefunction renormalization factor while the second, $M_r$, is a renormalized 
mass.  The continuum theory with no defects corresponds to $M_r=M=5$ and $Z=1$. In the presence of the defects, we find good agreement between the average points and a continuum Feynman propagator with $M_r=4.08$ and $Z=0.47$.  Interestingly, the effective mass of the scalar particle is shifted downward in the presence of defects.   In an interacting quantum field theory, shifts in the mass and wavefunction renormalization are a consequence of interactions, while in the present case, spacetime defects have a similar impact on the Feynman propagator.
\begin{figure}
\centering
 \includegraphics[width=0.5\textwidth]{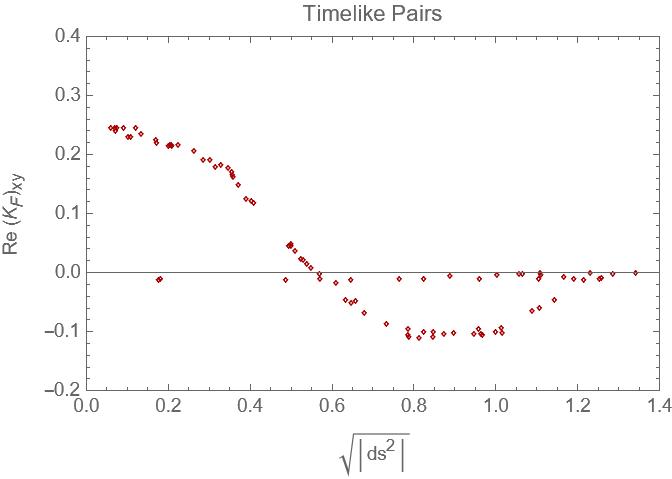}
\caption{Plot of the real part of the Feynman propagator $(K_F)_{ab}$ in Model A over the narrow timelike strip $|x| \leq 0.05$. The parameters are the same as those specified in Fig.~\ref{fig:two}. 
The spacetime point $v_b$ is chosen to be the point in this region (and within the unit causal diamond) with the largest $t$ coordinate.  This illustrates the discontinuities due to the defects in a plot 
that is single-valued.} \label{fig:three}
\end{figure}

We can avoid the multivaluedness of plotting $(K_F)_{ab}$ against the absolute value of the proper time separation between two 
points $v_a$ and $v_b$, by plotting the propagator over a thin slice of the sprinkling region and fixing $v_b$ so that the magnitude of 
the spacetime interval varies monotonically as one moves through the subset of points.   In Fig.~\ref{fig:three}, we consider the slice 
$|x| \leq 0.05$, where $x$ is the spatial coordinate, and fix $v_b$ to be the point within this region (and within the unit causal 
diamond) that has the largest $t$ coordinate.  By considering a narrow strip where the points $v_a$ have roughly the same spatial 
coordinate as $v_b$, but  different time coordinate, we preclude having more than one pair having the same magnitude of proper 
time separation.  Since most of the elements in the band will be ordinary spacetime points, we expect that the small number of 
defects in the band will show up as discontinuous jumps in the result. Fig.~\ref{fig:three} demonstrates this effect.
        
\subsection{Model B}
    
\begin{figure}[]
\centering
\subfloat[\label{subfig:4a}]{%
  \includegraphics[width=0.5\columnwidth]{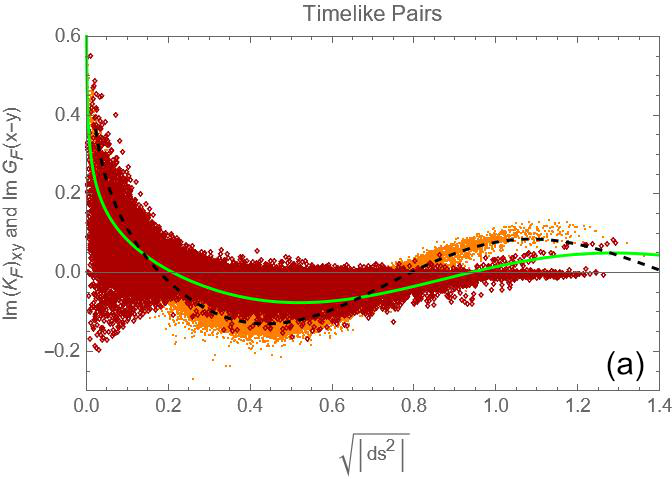}%
}\hfill
\subfloat[\label{subfig:4b}]{%
  \includegraphics[width=0.5\columnwidth]{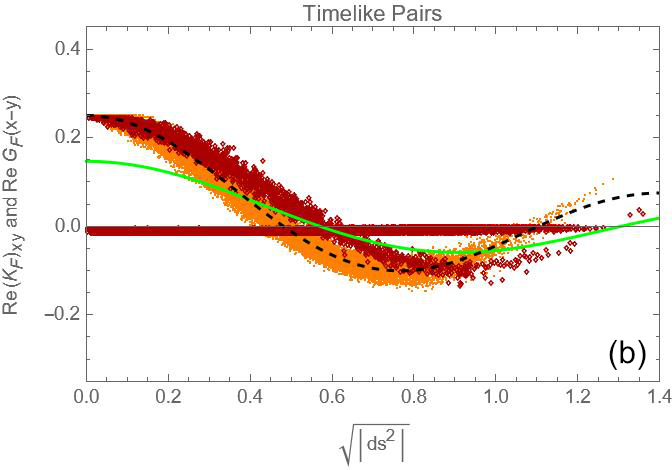}%
}\hfill
\subfloat[\label{subfig:4c}]{%
  \includegraphics[width=0.5\columnwidth]{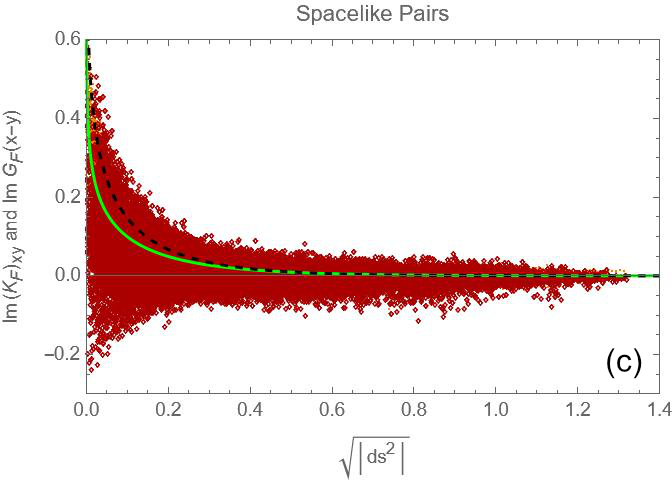}%
}
\subfloat[\label{subfig:4d}]{%
  \includegraphics[width=0.5\columnwidth]{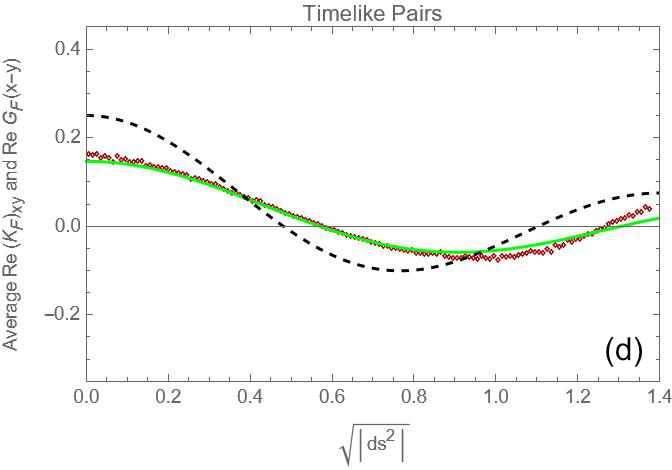}%
}
\caption{{\bf Model B} with $M=5$ and $\epsilon=1.0$. Feynman propagator for a 600-element causal set that includes 200 defect 
points in {\it spacelike} in-out pairs, in 1+1 dimensions.  The darker points (red in color) are the results; the lighter points (orange in color) are the results in the absence of defects.
The analogous plots shown for Model A in Fig.~\ref{fig:two} are displayed above.} \label{fig:four}
\end{figure}

This model would be identical to Model A, except that we now choose the in-defects and the out-defects to be spacelike separated. Thus, any path between spacelike separated elements must pass through at least one pair of in and out defects.  We obtain results
that are qualitatively similar to those of Model A, and we again provide a fit using the two-parameter function in Eq.~(\ref{eq:fitfnc}).
For Model B, we find $M_r=4.22$ and $Z=0.59$. Thus, the mass and wavefunction renormalization appear smaller in magnitude than the case with no defects present, much like model A. 

\begin{figure}
\centering
\subfloat[\label{subfig:5a}]{%
  \includegraphics[width=0.5\columnwidth]{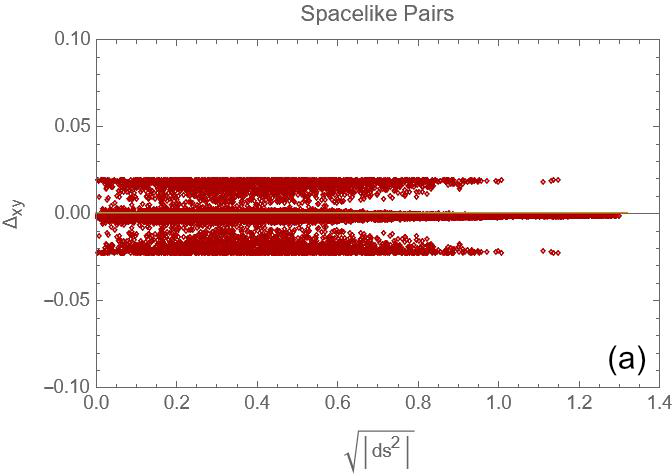}%
}\hfill
\subfloat[\label{subfig:5b}]{%
  \includegraphics[width=0.5\columnwidth]{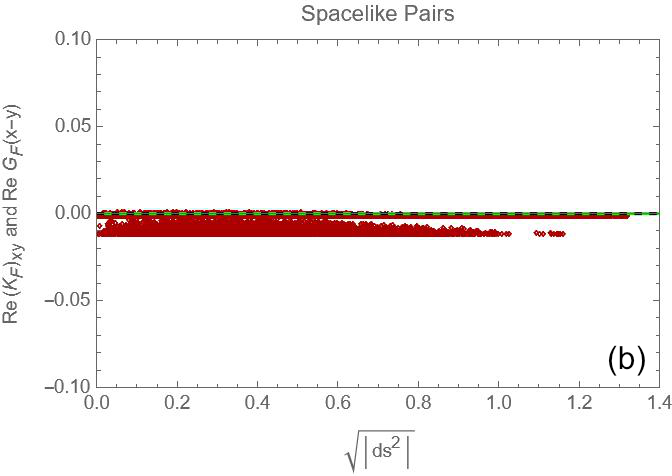}%
}
\caption{(a) Plot of the Pauli-Jordan matrix for spacelike separated pairs of elements in Model B, with the same parameter choices as 
in Fig.~\ref{fig:four}. The darker points (red in color) are the results; the lighter points (orange in color) are the results in the absence of 
defects. (b) Plot of the real part of the Feynman propagator for spacelike separated events.} \label{fig:five}
\end{figure}

Recall in Sec.~\ref{sec:models}, that the Pauli-Jordan matrix is related to the commutator of field operators at each causal set 
element, $[\phi_{x}, \phi_{y}]=\Delta_{xy}$. For all models that do not lead to a violation of causality, we expect the Pauli-Jordan 
function to vanish for spacelike-separated elements of the causal set, a fact that we have verified numerically in models A, C, D 
and E.  However, in Model B we have in-defects connected to out-defects with spacelike separation, which allows chains in the causal 
set to exist between spacelike separated events. Figure~\ref{subfig:5a} demonstrates that the Pauli-Jordan function in Model B is 
nonvanishing for points in the causal set that are spacelike separated, but vanishing when the number of defects are taken to zero.  Hence, we expect causality to be violated in Model B.

We also include the plot of the real part of the Feynman propagator for spacelike separated pairs in this model since it is non-trivial. In other models, where the interval between the in- and
out-defects was timelike, the real part of the propagator vanished for spacelike separated points in the causal set.   This can be understood by noting that chains between elements of the 
causal set that are spacelike separated are nonexistent in these models; hence, the elements of both $K_{R}$ and $K_{A}$ that correspond to spacelike separated points vanish, as do their combination in ${\rm Re} \, K_{F}=\frac{1}{2}(K_{R}+K_{A})$.
        
\subsection{Model C} 
    
\begin{figure}
\centering
\subfloat[\label{subfig:6a}]{%
  \includegraphics[width=0.5\columnwidth]{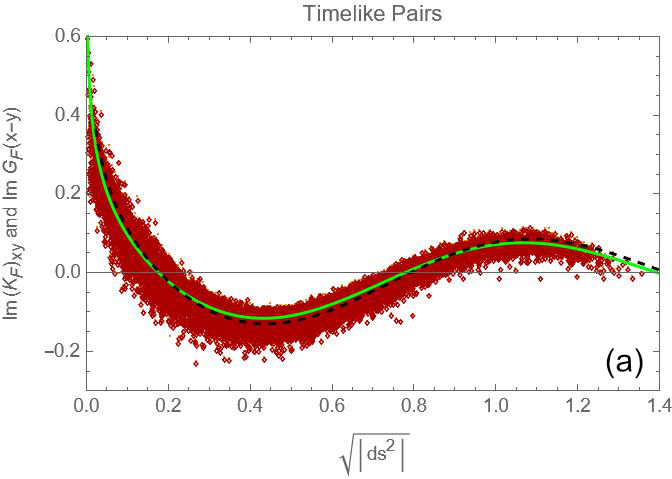}%
}\hfill
\subfloat[\label{subfig:6b}]{%
  \includegraphics[width=0.5\columnwidth]{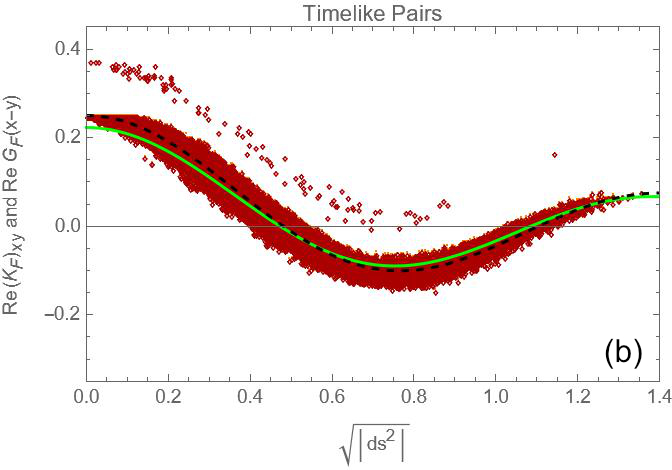}%
}\hfill
\subfloat[\label{subfig:6c}]{%
  \includegraphics[width=0.5\columnwidth]{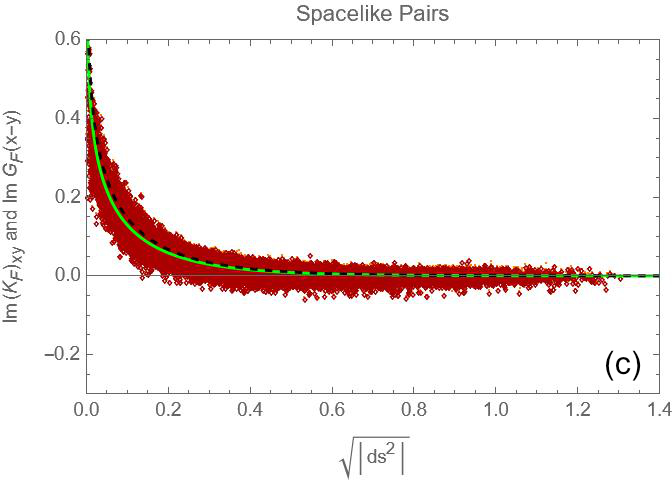}%
}
\subfloat[\label{subfig:6d}]{%
  \includegraphics[width=0.5\columnwidth]{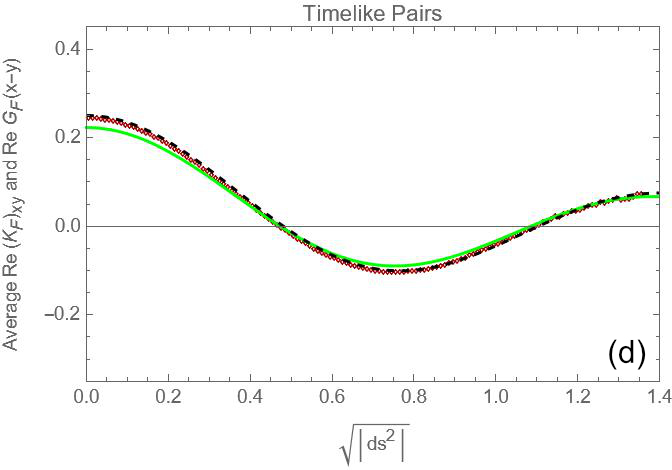}%
}
 \caption{{\bf Model C} with $M=5$ and $\epsilon=0.75$.  Feynman propagator for a 600-element causal set that includes 200 
 defect points in timelike in-out pairs, in 1+1 dimensions.  (a) and (b) show the imaginary and real parts of the propagator, 
 respectively, for points that are timelike separated; (c) shows the imaginary part for points that are spacelike separated (the real 
part vanishes).  The darker points (red in color) are the results; the lighter points (orange in color) are the results in the absence of defects.
The dashed curve (black in color) is the continuum propagator without defects and the solid curve (green in color) is the fitted curve to the 
average defective propagator; (d) is the same as (b) but only displays the average points in the presence of defects. \label{fig:six}}
\end{figure}

\begin{figure}
\centering
\subfloat[\label{subfig:7a}]{%
  \includegraphics[width=0.5\columnwidth]{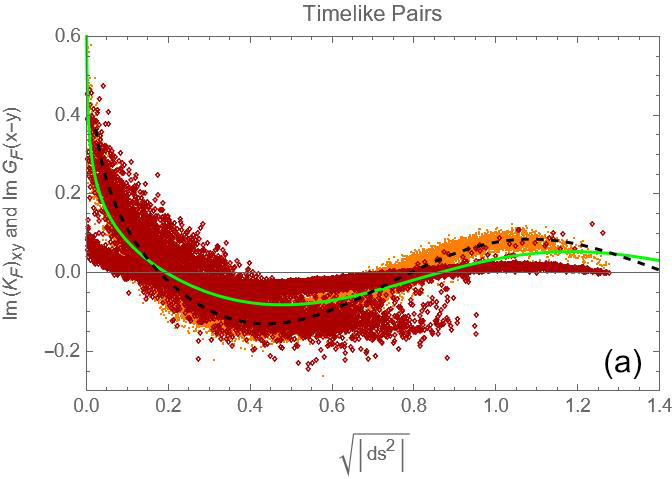}%
}\hfill
\subfloat[\label{subfig:7b}]{%
  \includegraphics[width=0.5\columnwidth]{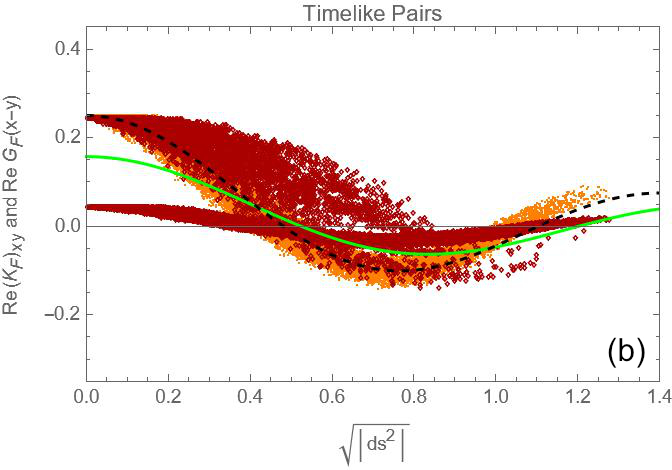}%
}\hfill
\subfloat[\label{subfig:7c}]{%
  \includegraphics[width=0.5\columnwidth]{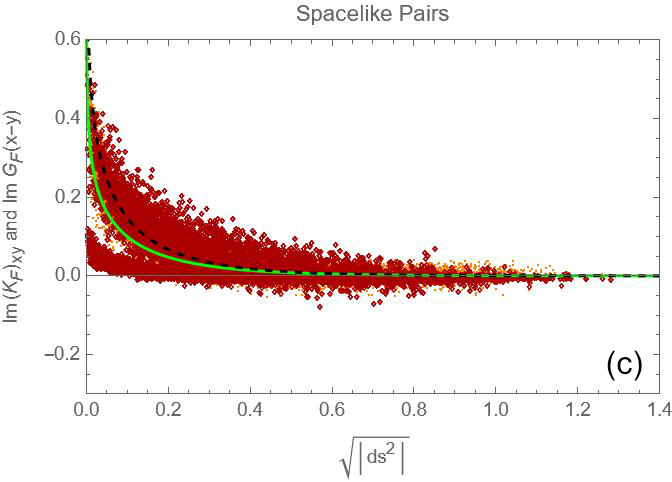}%
}
\subfloat[\label{subfig:7d}]{%
  \includegraphics[width=0.5\columnwidth]{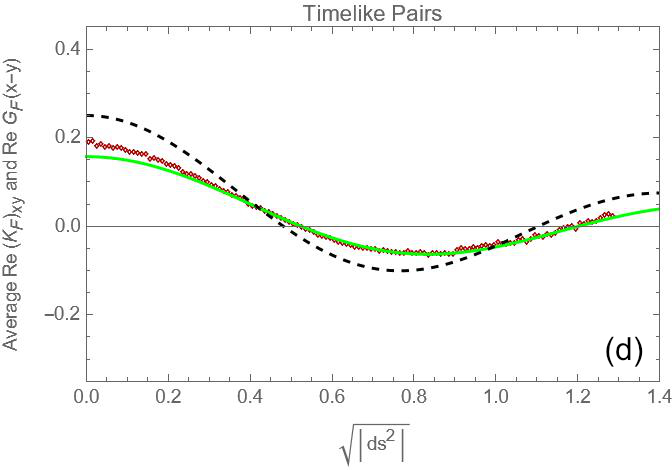}%
}
\caption{{\bf Model D} with $M=5$, $\xi=0.1$ and $\epsilon=0.5$.  Feynman propagator for a 500-element causal set that includes 
100 defect points, in 1+1 dimensions.  (a) and (b) show the imaginary and real parts of the propagator, respectively, for points that 
are timelike separated; (c) shows the imaginary part for points that are spacelike separated (the real part vanishes).   
The darker points (red in color) are the results; the lighter points (orange in color) are the results in the absence of defects.
The dashed curve (black in color) is the continuum propagator without defects and the solid curve (green in color) is the fitted curve to the 
average defective propagator; (d) is the same as (b) but only displays the average points in the presence of defects. \label{fig:seven}}
 \end{figure}
 
Model C relaxes the assumptions of Model A that in-defects can only beam to out-defects, and that no other trajectories can exit from 
in-defects or are incoming to out-defects.   Since all causal links between points are included, the defect-free limit can be obtained by 
either eliminating the defects, or setting the parameter $\epsilon=a =1/2$, whereby the defects behave no differently than ordinary 
points. One can see the dominant curves are not appreciably different than the case of the defect-free propagator, except in 
Fig.~\ref{subfig:6b} there is an additional sparse curve above the main branch that corresponds to chains connecting an in-defect to 
an out-defect.  Fitting the average propagator result using Eq.~(\ref{eq:fitfnc}), we find that $M_r=5.09$ and $Z=0.89$.   The fit is 
dominated by the main branch of the solution which has a significant contribution from propagation between pairs of ordinary spacetime 
points and where the effects of the defects for $\epsilon$ deviating from $0.5$ are small.

\begin{figure}
\centering
\subfloat[\label{subfig:8a}]{%
  \includegraphics[width=0.5\columnwidth]{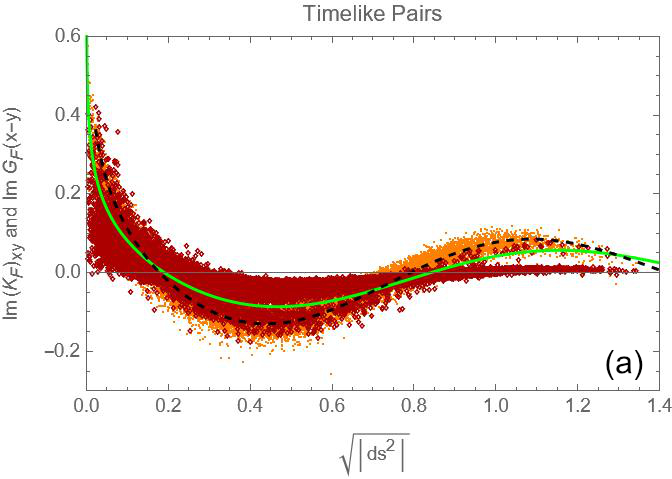}%
}\hfill
\subfloat[\label{subfig:8b}]{%
  \includegraphics[width=0.5\columnwidth]{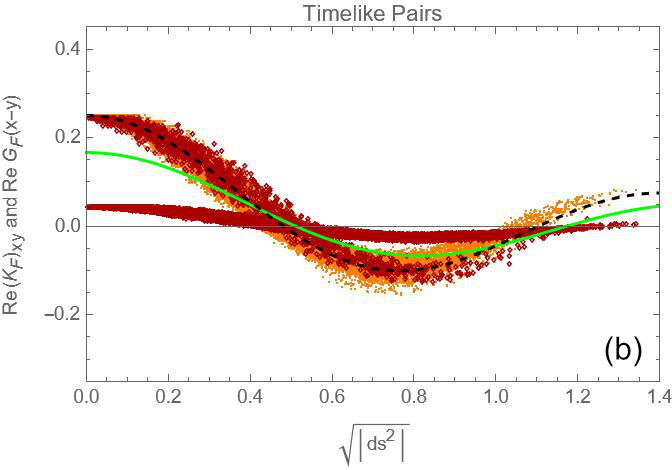}%
}\hfill
\subfloat[\label{subfig:8c}]{%
  \includegraphics[width=0.5\columnwidth]{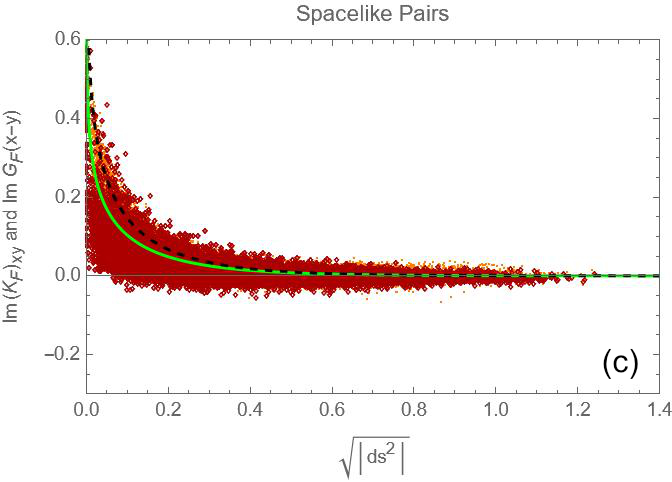}%
}
\subfloat[\label{subfig:8d}]{%
  \includegraphics[width=0.5\columnwidth]{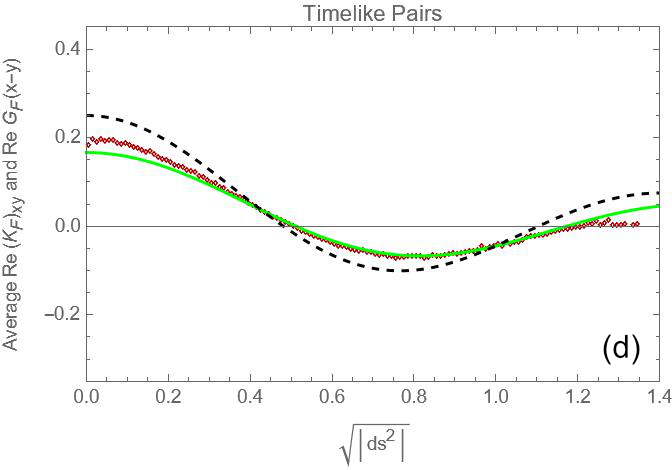}%
}
\caption{{\bf Model E} with $M=5$ and $\kappa=0.1$:  Feynman propagator for a 500-element causal set that includes 100 defect 
points, in 1+1 dimensions.  (a) and (b) show the imaginary and real parts of the propagator, respectively, for points that are timelike 
separated; (c) shows the imaginary part for points that are spacelike separated (the real part vanishes).  The darker points (red in color) 
are the results; the lighter points (orange in color) are the results in the absence of defects.  The dashed curve (black in color) is the continuum 
propagator without defects and the solid curve (green in color) is the fitted curve to the average defective propagator; (d) is the same as (b) but 
only displays the average points in the presence of defects. \label{fig:eight}}
\end{figure} 
\subsection{Model D}
    
In this model, it is easy to see at least two curves emerging that contribute to the average Feynman propagator.  The main branch
correspond to propagation between ordinary spacetime points while, for example, while the curve closest to the $x$-axis (most clearly
visible in Fig.~\ref{subfig:7b} represents propagation between ordinary points and defects, or vice versa. The fit of the average 
Feynman propagator to the propagator without defects in Fig. \ref{subfig:7d} gives $M_r=4.59$ and $Z=0.63$. Again, we see 
that the effect of the defects is to lower the mass of the particle and produce a smaller wavefunction renormalization.

\subsection{Model E}

The results of this model appear similar to model D which also has only one type of defect point that is distinguished by the 
weighting of the chains that include those points.  Based on Figs.~\ref{subfig:8a} and \ref{subfig:8b}, two curves dominate the 
contributions to the average propagator. The first one, which is most similar to the defect-free result, corresponds propagating from a 
spacetime point to any other point. The second, smaller curve, corresponds to propagating from a defect to any other point. The average 
points are fit well by Eq.~(\ref{eq:fitfnc}) with the parameters $M_r=4.70$ and $Z=0.67$. In this model, we again observe both a smaller 
effective mass and wavefunction renormalization.   

To conclude this section, we present in Table~\ref{table:one} the wavefunction renormalization and renormalized mass for each model, 
along with measures of the goodness of fit corresponding to subfigures (a), (b) and (c) in Figs.~\ref{fig:two}, \ref{fig:four}, \ref{fig:six}, \ref{fig:seven} and \ref{fig:eight}.

\begin{table}[h]
\begin{center}
\begin{tabular}{ccccccccccc}
\hline \hline
Model &\hspace{1em} &$M_r / M$ &\hspace{1em} & Z & \hspace{1em} &  $R^{2}_a$ & \hspace{1em} & $R^{2}_b$ & \hspace{1em} & $R^{2}_c$  \\ [0.5ex] 
\hline\hline
A & & 0.8151 & & 0.4650 & &  0.9214 & & 0.9474 & & 0.8104\\
B & &0.8445 & & 0.5859 & & 0.9487 & & 0.9775 & & 0.8865\\
C & &1.0171 & & 0.8911 & & 0.9765 & & 0.9883 & & 0.9786 \\
D & & 0.9178 & & 0.6284 & & 0.9784 & & 0.9727 & & 0.9348 \\
E & & 0.9409 & & 0.6653 & &  0.9692 & & 0.9728 & & 0.9325\\ [1ex]
\hline\hline 
\end{tabular} \vspace*{1em}
\caption{Parameter values for the fitting function, Eq.~(\ref{eq:fitfnc}), for Models A-E.  The last three columns shows the separate goodness of fit to the results 
shown in subfigures (a) through (c) of Figs.~\ref{fig:two}, \ref{fig:four}, \ref{fig:six}, \ref{fig:seven} and \ref{fig:eight}.} \label{table:one}
\end{center}
\end{table}

\section{Conclusions} \label{sec:conc} 
The difficulty of directly probing the Planck scale motivates considering the possibility of remnants of the physics of quantum gravity 
that might survive in the infrared. Spacetime defects are one such possibility, and their spacetime density (or the length scale given by 
the fourth root of its reciprocal in 3+1 dimensions) may be substantially sub-Planckian.   The possibility of nonlocal defects has been 
raised in the literature~\cite{Hossenfelder:2014hha,Hossenfelder:2013zda,Hossenfelder:2013yda}, but no compelling descriptions in the 
context of quantum field theory have been proposed.  The current work is an attempt to model the effects of nonlocal defects within 
quantum field theory, working within the framework of causal sets.

We have approached the problem by assuming that the effects of defects could be encoded by modifying the Green's functions used 
in the quantization of a field theory on a causal set.   In the case of a real scalar field, it has been shown in the literature on casual 
sets that these discrete Green's functions can be derived by summing chains or paths through causal set 
points~\cite{Johnston:2008za,Johnston:2009fr}.  Modifying the amplitude for particular particle chains that pass through defect points 
allowed us to define a number of scenarios.  For example, a defect that ``beams" an incident particle exclusively to a second 
specified defect point is one of the possibilities we consider; we refer the reader to Sec.~\ref{sec:models} for a summary of the others. 
Working in two-dimensional Minkowski spacetime, we numerically study in each case how the discrete propagator 
in position space differs from the  defect-free limit, where the latter is consistent, on average, with the continuum expectation as the 
number of points is taken large.   When defects are present, two points with the same invariant spacetime interval can have different 
propagation amplitudes, depending on whether the propagation is between two ordinary spacetime points, two defects, or a defect 
and an ordinary point.   In principle, with sufficient resolution, this breaking of translation invariance might be observable in scattering 
amplitudes.  Otherwise, the average effect is to deform the propagator one would extract in a continuum effective field theory.  By fitting our results, we find that this modification is well described by a defect-induced mass and wavefunction renormalization.   Shifts in particle masses due to defects have also been discussed previously in Ref.~\cite{Klinkhamer:2017nhl}.

Admittedly, the present work is far from realistic.  We consider only a real scalar field and work in two spacetime dimensions, where 
numerical study is tractable.  We focus only on the two-point function in position space.   The value of this first study is that it gives a 
concrete framework for modeling the physics of nonlocal defects that might be generalized to more realistic theories, and in a more 
realistic number of spacetime dimensions.  It is also possible that other approaches to defining quantum field theories on causal sets 
may lead to better formulations of the problem.   It is our hope that the present work will stimulate additional work in these directions.

%%%%%%%%%%%%%%%%%%%%%%%%%%%%%%%%%%%%%%%%%%%%%%%%%%%%%%%%%%%
\begin{acknowledgments} 
We thank the NSF for support under Grant PHY-2112460.  
\end{acknowledgments}

%\appendix


\begin{thebibliography}{99}

%\cite{Addazi:2021xuf}
\bibitem{Addazi:2021xuf}
A.~Addazi, J.~Alvarez-Muniz, R.~Alves Batista, G.~Amelino-Camelia, V.~Antonelli, M.~Arzano, M.~Asorey, J.~L.~Atteia, S.~Bahamonde and F.~Bajardi, \textit{et al.}
``Quantum gravity phenomenology at the dawn of the multi-messenger era\textemdash{}A review,''
Prog. Part. Nucl. Phys. \textbf{125}, 103948 (2022)
%doi:10.1016/j.ppnp.2022.103948
\href{https://arxiv.org/abs/2111.05659}{arXiv:2111.05659 [hep-ph]}.
%240 citations counted in INSPIRE as of 25 Sep 2023

%\cite{Hossenfelder:2014hha}
\bibitem{Hossenfelder:2014hha}
S.~Hossenfelder,
``Theory and Phenomenology of Spacetime Defects,''
Adv. High Energy Phys. \textbf{2014}, 950672 (2014)
%doi:10.1155/2014/950672
\href{https://arxiv.org/abs/1401.0276}{arXiv:1401.0276 [hep-ph]}.
%20 citations counted in INSPIRE as of 15 Sep 2023

%\cite{Hossenfelder:2013zda}
\bibitem{Hossenfelder:2013zda}
S.~Hossenfelder,
``Phenomenology of Space-time Imperfection II: Local Defects,''
Phys. Rev. D \textbf{88}, no. 12, 124031 (2013)
%doi:10.1103/PhysRevD.88.124031
\href{https://arxiv.org/abs/1309.0314}{arXiv:1309.0314 [hep-ph]}.
%13 citations counted in INSPIRE as of 15 Sep 2023

%\cite{Hossenfelder:2013yda}
\bibitem{Hossenfelder:2013yda}
S.~Hossenfelder,
``Phenomenology of Space-time Imperfection I: Nonlocal Defects,''
Phys. Rev. D \textbf{88}, no. 12, 124030 (2013)
%doi:10.1103/PhysRevD.88.124030
\href{https://arxiv.org/abs/1309.0311}{arXiv:1309.0311 [hep-ph]}.
%13 citations counted in INSPIRE as of 15 Sep 2023

%\cite{Schreck:2012pf}
\bibitem{Schreck:2012pf}
M.~Schreck, F.~Sorba and S.~Thambyahpillai,
``Simple model of pointlike spacetime defects and implications for photon propagation,''
Phys. Rev. D \textbf{88}, no. 12, 125011 (2013)
%doi:10.1103/PhysRevD.88.125011
\href{https://arxiv.org/abs/1211.0084}{arXiv:1211.0084 [hep-th]}.
%8 citations counted in INSPIRE as of 04 Oct 2023

%\cite{Klinkhamer:2017nhl}
\bibitem{Klinkhamer:2017nhl}
F.~R.~Klinkhamer and J.~M.~Queiruga,
``Mass generation by a Lorentz-invariant gas of spacetime defects,''
Phys. Rev. D \textbf{96}, no. 7, 076007 (2017)
%doi:10.1103/PhysRevD.96.076007
\href{https://arxiv.org/abs/1703.10585}{arXiv:1703.10585 [hep-th]}.
%4 citations counted in INSPIRE as of 20 Sep 2023

%\cite{Queiruga:2017nfe}
\bibitem{Queiruga:2017nfe}
J.~M.~Queiruga,
``Particle propagation on spacetime manifolds with static defects,''
J. Phys. A \textbf{51}, no. 4, 045401 (2018)
%doi:10.1088/1751-8121/aa9b51
\href{https://arxiv.org/abs/1703.03606}{arXiv:1703.03606 [hep-th]}.
%2 citations counted in INSPIRE as of 04 Oct 2023

%\cite{Bombelli:1987aa}
\bibitem{Bombelli:1987aa}
L.~Bombelli, J.~Lee, D.~Meyer and R.~Sorkin,
``Space-Time as a Causal Set,''
Phys. Rev. Lett. \textbf{59}, 521-524 (1987)
%doi:10.1103/PhysRevLett.59.521
%675 citations counted in INSPIRE as of 26 Sep 2023

%\cite{Surya:2019ndm}
\bibitem{Surya:2019ndm}
For a review, see S.~Surya,
``The causal set approach to quantum gravity,''
Living Rev. Rel. \textbf{22}, no.~1, 5 (2019)
%doi:10.1007/s41114-019-0023-1
\href{https://arxiv.org/abs/1903.11544}{arXiv:1903.11544 [gr-qc]}.
%121 citations counted in INSPIRE as of 26 Sep 2023

%\cite{Johnston:2008za}
\bibitem{Johnston:2008za} 
S.~Johnston,
``Particle propagators on discrete spacetime,''
Class. Quant. Grav. \textbf{25}, 202001 (2008)
%doi:10.1088/0264-9381/25/20/202001
\href{https://arxiv.org/abs/0806.3083}{arXiv:0806.3083 [hep-th]}.
%61 citations counted in INSPIRE as of 13 Sep 2023

%\cite{Johnston:2009fr}
\bibitem{Johnston:2009fr}
S.~Johnston,
``Feynman Propagator for a Free Scalar Field on a Causal Set,''
Phys. Rev. Lett. \textbf{103}, 180401 (2009)
%doi:10.1103/PhysRevLett.103.180401
\href{https://arxiv.org/abs/0909.0944}{arXiv:0909.0944 [hep-th]}.
%64 citations counted in INSPIRE as of 13 Sep 2023

%\cite{Shuman:2023agb}
\bibitem{Shuman:2023agb}
S.~Shuman,
``Path Sums for Propagators in Causal Sets,''
\href{https://arxiv.org/abs/2307.08864}{arXiv:2307.08864 [gr-qc]}.
%0 citations counted in INSPIRE as of 01 Oct 2023

%\cite{Dowker:2003hb}
\bibitem{Dowker:2003hb}
F.~Dowker, J.~Henson and R.~D.~Sorkin,  
``Quantum gravity phenomenology, Lorentz invariance and discreteness,''
Mod. Phys. Lett. A \textbf{19}, 1829-1840 (2004)
%doi:10.1142/S0217732304015026
\href{https://arxiv.org/abs/gr-qc/0311055}{arXiv:gr-qc/0311055 [gr-qc]}.
%133 citations counted in INSPIRE as of 04 Oct 2023

%\cite{Bombelli:2006nm}
\bibitem{Bombelli:2006nm}
L.~Bombelli, J.~Henson and R.~D.~Sorkin,
``Discreteness without symmetry breaking: A Theorem,''
Mod. Phys. Lett. A \textbf{24}, 2579-2587 (2009)
%doi:10.1142/S0217732309031958
\href{https://arxiv.org/abs/gr-qc/0605006}{arXiv:gr-qc/0605006 [gr-qc]}.
%98 citations counted in INSPIRE as of 04 Oct 2023

%\cite{Sorkin:2007qi}
\bibitem{Sorkin:2007qi}
R.~D.~Sorkin,
``Does locality fail at intermediate length-scales,''
\href{https://arxiv.org/abs/gr-qc/0703099}{arXiv:gr-qc/0703099 [gr-qc]}.
%67 citations counted in INSPIRE as of 01 Oct 2023

\end{thebibliography}
\end{document}